\documentclass[aps,preprint]{revtex4}
\usepackage{amsfonts}
\usepackage{amsmath}
\usepackage{amssymb}
\usepackage{graphicx}

\input{tcilatex}
\begin{document}

\preprint{}
\title[ ]{ Restricted Class of Colliding Einstein-Yang-Mills Plane Waves}
\author{Ozay Gurtug}
\email{ozay.gurtug@emu.edu.tr}
\affiliation{Department of Physics, Eastern Mediterranean University, G. Magusa, north
Cyprus, via Mersin 10, Turkey.}
\author{Mustafa Halilsoy}
\email{mustafa.halilsoy@emu.edu.tr}
\affiliation{Department of Physics, Eastern Mediterranean University, G. Magusa, north
Cyprus, via Mersin 10, Turkey.}

\begin{abstract}
By gauging an Abelian electromagnetic (em) solution through a non-Abelian
transformation and in accordance with a theorem proved long time ago, we
construct a simple class of colliding Einstein-Yang-Mills (EYM) plane waves.
The solution is isometric to the Wu-Yang charged Kerr-Newman (KN) black hole
and shares much of the properties satisfied by colliding Einstein-Maxwell
(EM) plane waves. In the linear polarization limit with unit degenerate
charge it reduces to the Bell-Szekeres (BS) solution for colliding em shock
waves.
\end{abstract}

\maketitle

\section{Introduction}

Starting with 1970s, through 80s, were the years in which the subject of
colliding waves in general relativity attracted much interest. All known
results were compiled later into a cataloque of solutions\cite{JB}. In later
decades the interest in the subject continued with less momentum,
concentrating more on the sophisticated fields such as dilaton, axion and
torsion coupled \ to gravity and electromagnetic (em) fields. Formation of
Cauchy horizon / singularity, and under what conditions the horizon remains
stable dominated most discussions to date. To our knowledge the discussion
has not been conclusive to the satisfaction of all yet. Our aim in this
paper is not to contribute in this particular direction, but rather to point
out a restricted class of colliding Yang-Mills (YM) waves which behaves
almost em-like.

The motivating factor to consider such a problem anew relies on the recent
attempts of Large Hadron Collisions (LHC) at TeV scale that maintain the
prime agenda at CERN. As the protons (anti-protons) are boosted to almost
the speed of light they behave more wave-like than particle-like whose
collisions are reminiscent of waves colliding in general relativity. Since
the inner/color structure of hadrons have constituent YM fields, collision
of such waves deserves further investigation.

Within this context although there is a large collection of Einstein-Maxwell
(EM) solutions available in the literature \cite{JB}\cite{MH}, extension of
the problem to the Einstein-Yang-Mills (EYM) remained ever open. We aim to
contribute in this regard, at least partly, for the gauge group $SO(3)$ and
pave the way for further solutions underlying various gauge groups. Our
starting point is a Theorem proved long time ago by P. Yasskin \cite{PY} in
connection with Yang-Mills (YM) fields in a curved spacetime. The method in
Yasskin's Theorem is to start from \ an Abelian $U(1)$ em solution and map
it through a non-Abelian transformation to the YM problem. In this process,
naturally, the degenerate YM charges are defined from the original em
charge. By employing the Wu-Yang ansatz \cite{WY} for the YM fields in the
trapped region of the Kerr-Newman (KN) black hole we construct solution that
describes colliding EYM plane waves. In this process, rotation of the black
hole transforms into the cross polarization of the colliding waves. In the
linear polarization limit we recover the colliding em wave solution due to
Bell and Szekeres (BS)\cite{BS}. \ More generally, any colliding EM
metric/field can be shown to represent at the same time colliding EYM waves,
in a restricted sense provided the YM field is defined according to the
Yasskin's Theorem. This guarantees that the incoming YM fields are em-like
plane waves so that they do not create extra currents. It can be anticipated
that non-planar waves will induce their own sources through self-interaction
which is the case that we avoid in the present study. Although YM fields is
known to be a subject in the realm of quantum (chromodynamics), our
treatment is entirely classical here. Stated otherwise, we adopt the
viewpoint that anything that interacts with the classical gravity must
itself behave classical.

Organization of the paper is as follows. In section II, we introduce YM
fields in a KN black hole geometry. Colliding EYM plane waves follows in
section III. Section IV, concentrates on the linear polarization of the
waves. We complete the paper with conclusion in section V.

\section{KN Black Hole and YM Fields.}

The KN black hole solution is given by the line element

\begin{equation}
ds^{2}=\frac{U^{2}}{\rho ^{2}}\left( dt-\overline{a}\sin ^{2}\theta d\varphi
\right) ^{2}-\frac{\sin ^{2}\theta }{\rho ^{2}}\left[ Fd\varphi -\overline{a}%
dt\right] ^{2}-\frac{\rho ^{2}}{U^{2}}dr^{2}-\rho ^{2}d\theta ^{2},
\end{equation}%
where

\begin{equation}
U^{2}=r^{2}-2mr+\overline{a}^{2}+Q^{2},\text{ \ \ \ \ \ \ }\rho ^{2}=r^{2}+%
\overline{a}^{2}\cos ^{2}\theta ,\text{ \ \ \ \ \ }F=r^{2}+\overline{a}^{2},%
\text{\ }
\end{equation}%
in which $\overline{a}$ is the parameter of rotation and $Q$ is the em
charge. By invoking Yasskin's Theorem \cite{PY} we extend this solution to
represent a particular YM field as follows. A suitable YM gauge potential
1-form $A^{i}=A_{\mu }^{i}dx^{\mu },\left( i=1,2,3\right) $ is chosen as

\begin{equation}
A^{i}=\frac{1}{\rho ^{2}}Q^{i}\cos \theta \left[ \left( r^{2}+\overline{a}%
^{2}\right) d\varphi -\overline{a}dt\right] .
\end{equation}%
Here the gauge charge $Q^{i}$ satisfies the constraint

\begin{equation}
\gamma _{ij}Q^{i}Q^{j}=Q^{2},
\end{equation}%
with the invariant group metric $\gamma _{ij}=\delta _{ij}$ and $Q$ is the
charge of the KN \ black hole. The YM field 2-form is defined by $F^{i}=%
\frac{1}{2}F_{\mu \nu }^{i}dx^{\mu }\wedge dx^{\nu },$ where $\wedge $
stands for the wedge product and

\begin{equation}
F_{\mu \nu }^{i}=\partial _{\mu }A_{\nu }^{i}-\partial _{\nu }A_{\mu }^{i}+%
\frac{1}{2Q}\epsilon _{jk}^{i}A_{\mu }^{j}A_{\nu }^{k},
\end{equation}%
in which $\epsilon _{jk}^{i}$ is the structure constant for the group. In
the $SO(3)$ basis, $T^{i}$ $\left( i=1,2,3\right) $ are given by

\begin{equation}
T^{1}=\left( 
\begin{array}{ccc}
0 & 0 & 0 \\ 
0 & 0 & -1 \\ 
0 & 1 & 0%
\end{array}%
\right) ,\text{ \ \ \ }T^{2}=\left( 
\begin{array}{ccc}
0 & 0 & 1 \\ 
0 & 0 & 0 \\ 
-1 & 0 & 0%
\end{array}%
\right) ,\text{ \ \ \ }T^{3}=\left( 
\begin{array}{ccc}
0 & -1 & 0 \\ 
1 & 0 & 0 \\ 
0 & 0 & 0%
\end{array}%
\right) ,
\end{equation}%
and the YM potential 1-form has the representation

\begin{equation}
A=A_{\mu }^{i}T^{i}dx^{\mu }.
\end{equation}%
For the particular choice ($Q^{3}=Q\neq 0=Q^{1}=Q^{2}$) we have

\begin{eqnarray}
A_{\varphi }^{3} &=&\frac{Q\cos \theta }{\rho ^{2}}\left( r^{2}+\overline{a}%
^{2}\right) , \\
A_{t}^{3} &=&-\frac{\overline{a}Q\cos \theta }{\rho ^{2}},\text{ \ \ \ \ \ \
\ }  \notag
\end{eqnarray}%
which has both electric and magnetic components. If $\overline{a}=0,$ i.e.
for the Reissner-Nordstrom black hole, we have a pure magnetic field. Now we
apply a gauge transformation on $A_{\mu }$ through

\begin{equation}
A_{\mu }\rightarrow \widetilde{A}_{\mu }=GA_{\mu }G^{-1}-Q\left( \partial
_{\mu }G\right) G^{-1},
\end{equation}%
where the gauge transformation matrix $G$ is

\begin{equation}
G=\left( 
\begin{array}{ccc}
\sin \varphi & \cos \varphi \cos \theta & \cos \varphi \sin \theta \\ 
-\cos \varphi & \sin \varphi \cos \theta & \sin \varphi \sin \theta \\ 
0 & -\sin \theta & \cos \theta%
\end{array}%
\right) .
\end{equation}%
We obtain the components of the new gauge potential 1-forms (after
suppressing the tilde over $A_{\mu }$ ) as

\begin{eqnarray}
A^{1} &=&\frac{Q}{2\rho ^{2}}\sin 2\theta \cos \varphi \left[ \left( r^{2}+%
\overline{a}^{2}\right) d\varphi -\overline{a}dt\right] +Q\sin \varphi
d\theta , \\
A^{2} &=&\frac{Q}{2\rho ^{2}}\sin 2\theta \sin \varphi \left[ \left( r^{2}+%
\overline{a}^{2}\right) d\varphi -\overline{a}dt\right] -Q\cos \varphi
d\theta ,  \notag \\
A^{3} &=&-\frac{Q}{\rho ^{2}}(r^{2}\sin ^{2}\theta d\varphi +\overline{a}%
\cos ^{2}\theta dt).  \notag
\end{eqnarray}%
In the next section we shall transform these potentials and metric (1) into
the colliding wave spacetime and interpret it as a solution to the colliding
EYM problem.

\section{Colliding EYM Plane Waves.}

The metric function $U^{2}=r^{2}-2mr+\overline{a}^{2}+Q^{2}=0,$ in Eq.(2)
has two roots, $r=r_{+}$ (the outer horizon) and $,$ $r=r_{-}$ (the inner
horizon). By the particular choice of parameters it is possible to make $%
r_{+}=r_{-}$ , which is called the extremal case, however, we shall consider
here only the case $r_{+}>r_{-}$ $\neq 0.$ It can easily be seen that for \ $%
r_{-}<r<r_{+}$, $U^{2}<0,$ which makes the spacetime to admit two spacelike
Killing vectors apt for colliding waves. For simplicity we choose the mass, $%
m=1$\ and apply the following transformation

\begin{eqnarray}
r &=&1+\alpha \tau ,\text{ \ \ \ }\sigma =\cos \theta ,\text{ \ \ \ }%
t=\alpha x,\text{ \ \ \ \ }y=\varphi , \\
&&\left( \alpha =\sqrt{1-\overline{a}^{2}-Q^{2}}=\frac{1}{p}\right)  \notag
\end{eqnarray}%
to the line element (1). After an overall scaling of the metric we obtain

\begin{equation}
ds^{2}=X\left( \frac{d\tau ^{2}}{\Delta }-\frac{d\sigma ^{2}}{\delta }%
\right) -X^{-1}\left( Rdx^{2}+Edy^{2}-2Gdxdy\right) ,
\end{equation}%
where

\begin{eqnarray}
X &=&\left( p+\tau \right) ^{2}+a_{0}^{2}\sigma ^{2},\text{ \ \ \ \ \ }%
R=\Delta +a_{0}^{2}\delta ,\text{\ \ \ \ \ \ } \\
E &=&\Delta A^{2}+\delta B^{2},\text{ \ \ \ \ \ \ }G=\Delta A+a_{0}\delta B,
\notag \\
A &=&a_{0}\delta ,\text{ \ \ \ \ \ \ }B=\left( p+\tau \right) ^{2}+a_{0}^{2},
\notag \\
\Delta &=&1-\tau ^{2},\text{ \ \ \ \ \ \ }\delta =1-\sigma ^{2},  \notag
\end{eqnarray}%
in which

\begin{equation}
\tau =\sin \left( au\theta \left( u\right) +bv\theta \left( v\right) \right)
,\text{ \ \ \ \ \ }\sigma =\sin \left( au\theta \left( u\right) -bv\theta
\left( v\right) \right) ,
\end{equation}%
are the new coordinates in terms of the null \ coordinates $\left(
u,v\right) $ and $\left( a,b\right) $ are arbitrary constants. We note also
that $a_{0}=\overline{a}p.$ By introducing another constant $q=Qp$, we can
check that the constraint condition

\begin{equation}
p^{2}-a_{0}^{2}-q^{2}=1,
\end{equation}%
holds. The YM potential 1-forms are now

\begin{eqnarray}
A^{1} &=&\frac{Q\sigma \sqrt{\delta }}{X}\cos y\left( Bdy-a_{0}dx\right)
-Q\sin y\frac{d\sigma }{\sqrt{\delta }}, \\
A^{2} &=&\frac{Q\sigma \sqrt{\delta }}{X}\sin y\left( Bdy-a_{0}dx\right)
+Q\cos y\frac{d\sigma }{\sqrt{\delta }},  \notag \\
A^{3} &=&\frac{Q\sigma ^{2}}{X}\left( Bdy-a_{0}dx\right) -dy.  \notag
\end{eqnarray}%
We note that in the null coordinates $\left( u,v\right) $ we have

\begin{eqnarray}
d\sigma &=&-\sqrt{\delta }\left( a\theta \left( u\right) du-b\theta \left(
v\right) dv\right) , \\
d\tau &=&\sqrt{\Delta }\left( a\theta \left( u\right) du+b\theta \left(
v\right) dv\right) ,  \notag
\end{eqnarray}%
in which $\theta \left( u\right) $ and $\theta \left( v\right) $ are the
unit step functions introduced as the requirement of the collision problem.
Let us note that insertion of the step functions must be checked critically
to ensure the absence of any redundant current sheets. In most cases such an
insertion fails to work but here, due to its em analogy it does work. The YM
field 2-form $F^{i}$ are given by.

\begin{equation*}
F^{i}=Q\sqrt{\delta }\left( \cos y,\sin y,\frac{\sigma }{\sqrt{\delta }}%
\right) F,
\end{equation*}%
where the 2-form $F$ is

\begin{equation}
F=\frac{1}{X^{2}}\left[ \left( \left( p+\tau \right) ^{2}-a_{0}^{2}\sigma
^{2}\right) d\sigma -2\sigma \left( p+\tau \right) d\tau \right] \wedge
\left( Bdy-a_{0}dx\right) ,
\end{equation}%
while its dual takes the form

\begin{equation}
^{\ast }F=\frac{1}{X^{2}}\left[ \left( \left( p+\tau \right)
^{2}-a_{0}^{2}\sigma ^{2}\right) d\tau \wedge \left( dx-a_{0}\delta
dy\right) -2a_{0}\sigma \left( p+\tau \right) d\sigma \wedge \left(
Bdy-a_{0}dx\right) \right] .
\end{equation}

Without much difficulty, which is more transparent in the $\left( \tau
,\sigma ,x,y\right) $, compared with the $\left( u,v,x,y\right) $
coordinates, one can show that the integrability equations

\begin{equation}
dF+A\wedge F=0,
\end{equation}%
and the YM equations

\begin{equation}
d^{\ast }F+A\wedge ^{\ast }F=0,
\end{equation}%
are all satisfied. This solves the problem of colliding EYM plane waves
where the YM field is obtained from the theorem of Yasskin while the metric
is the inner horizon region of the KN geometry explored first within the
context of EM by Chandrasekhar and Xanthopoulos \cite{CX2}. For a detailed
analysis of this spacetime we refer to \cite{CX2}.

\section{The Linear Polarization Limit.}

In this section we consider the metric obtained in the previous section and
set $a_{0}=0$ to make waves linearly polarized. In the null coordinates
after an overall scaling the line element takes the form

\begin{equation}
ds^{2}=\Sigma ^{2}\left( 2dudv-\delta dy^{2}\right) -\frac{\Delta }{\Sigma
^{2}}dx^{2},
\end{equation}%
where

\begin{equation}
\Sigma =1+\alpha \tau ,\ \ \ \ \Delta =1-\tau ^{2},\ \ \ \ \ \delta
=1-\sigma ^{2},\ \ \ \ \ \alpha =\sqrt{1-Q^{2}},
\end{equation}%
while the $SO(3)$ valued gauge potential 1-forms are

\begin{eqnarray}
A^{1} &=&Q\left[ \left( \sigma \sqrt{\delta }\cos y\right) dy-\left( \sin
y\right) \left( a\theta \left( u\right) du-b\theta \left( v\right) dv\right) %
\right] , \\
A^{2} &=&Q\left[ \left( \sigma \sqrt{\delta }\sin y\right) dy+\left( \cos
y\right) \left( a\theta \left( u\right) du-b\theta \left( v\right) dv\right) %
\right] ,  \notag \\
A^{3} &=&-Q\delta dy.  \notag
\end{eqnarray}

The YM field 2-form $F^{i}$ and $^{\ast }F^{i}$can be expressed as

\begin{eqnarray}
F^{i} &=&Q\delta \left( \cos y,\sin y,\frac{\sigma }{\sqrt{\delta }}\right) %
\left[ a\theta \left( u\right) du-b\theta \left( v\right) dv\right] \wedge
dy, \\
^{\ast }F^{i} &=&\frac{Q\sqrt{\Delta \delta }}{\left( \Sigma \right) ^{2}}%
\left( \cos y,\sin y,\frac{\sigma }{\sqrt{\delta }}\right) \left[ a\theta
\left( u\right) du+b\theta \left( v\right) dv\right] \wedge dx  \notag
\end{eqnarray}

In the null tetrad basis 1-forms $\left( l,n,m,\overline{m}\right) $ of
Newman and Penrose

\begin{eqnarray}
l &=&\Sigma du, \\
n &=&\Sigma dv,  \notag \\
m+\overline{m} &=&\sqrt{2}\Sigma \sqrt{\delta }dy,  \notag \\
m-\overline{m} &=&\sqrt{2}i\frac{\sqrt{\Delta }}{\Sigma }dx,  \notag
\end{eqnarray}%
the energy-momentum tensor $T_{\mu \nu }$ becomes,

\begin{equation}
4\pi T_{\mu \nu }=\frac{Q^{2}}{\Sigma ^{4}}\left[ a^{2}\theta \left(
u\right) l_{\mu }l_{\nu }+b^{2}\theta \left( v\right) n_{\mu }n_{\nu
}+ab\theta \left( u\right) \theta \left( v\right) \left( m_{\mu }m_{\nu }+%
\overline{m}_{\mu }\overline{m}_{\nu }\right) \right] .
\end{equation}%
Prior to the collision the incoming, coupled EYM plane waves are obtained by
setting $v<0$ \ $(u<0)$ in Eq.(24). For $u<0$ and $v<0$ we have a flat space
in which the YM field vanishes. For $u>0$ and $v>0$ , the Weyl scalars $\Psi
_{2},$ $\Psi _{0}$ and $\Psi _{4}$ are all regular as can be checked from

\begin{eqnarray}
\Psi _{2} &=&\frac{\alpha \left( \alpha +\tau \right) }{\Sigma ^{4}}ab\theta
\left( u\right) \theta \left( v\right) , \\
\Psi _{4} &=&-\frac{a\delta \left( u\right) \left[ \alpha +\sin \left(
bv\theta \left( v\right) \right) \right] }{\cos \left( bv\theta \left(
v\right) \right) \left[ 1+\alpha \sin \left( bv\theta \left( v\right)
\right) \right] }+3a^{2}\theta \left( u\right) \frac{\alpha \left( \alpha
+\tau \right) }{\Sigma ^{4}},  \notag \\
\Psi _{0} &=&-\frac{b\delta \left( v\right) \left[ \alpha +\sin \left(
au\theta \left( u\right) \right) \right] }{\cos \left( au\theta \left(
u\right) \right) \left[ 1+\alpha \sin \left( au\theta \left( u\right)
\right) \right] }+3b^{2}\theta \left( v\right) \frac{\alpha \left( \alpha
+\tau \right) }{\Sigma ^{4}},  \notag
\end{eqnarray}%
satisfying the type-D condition $\Psi _{0}\Psi _{4}=9\Psi _{2}^{2}.$ The
invariants $I=2\left( \Psi _{0}\Psi _{4}+3\Psi _{2}^{2}\right) $ and $%
J=6\Psi _{2}\left( \Psi _{0}\Psi _{4}-\Psi _{2}^{2}\right) $ are both
regular as they should be in the inside of the collision region. On the
boudaries $u=0,bv=\pi /2$ $(v=0,au=\pi /2)$, however, \ there are null
singularities which are also present in the problem of colliding em shock
waves\cite{BS}. The pathology involved ( if any) on the hypersurfaces $\tau
=1$ and $\sigma =\pm 1$ has been discussed extensively in the literature of
colliding waves ( see Ref.\cite{JB} and references therein). An analytic
extension beyond the horizon $(\tau =1)$, albeit it is a non - unique
process, reveals the geodesics completeness and other issues \cite{GH}.

The metric (23) \ has well known limits. For $Q=0$ $\left( \alpha =1\right) $
the YM field vanishes and one recovers the colliding gravitational waves
locally isometric to the Schwarzschild interior\cite{JB}. For $Q=1$ $\left(
\alpha =0\right) $ we have the case of colliding em waves \cite{BS}. This
implies that the Bell-Szekeres metric can be interpreted at the same time to
represent colliding YM plane waves for a unit charge $Q=1$.

The Wu-Yang ansatz solution\cite{WY} for the static YM fields has the gauge
potential

\begin{equation}
A^{i}=-\frac{Q}{r^{2}}\epsilon _{jk}^{i}x^{k}dx^{j}
\end{equation}%
where $x^{i}$ stands for the Cartesian coordinates and $Q=Q^{3}\neq 0,$ is
the only non-zero gauge charge. By the substitution, $x^{1}=r\sin \theta
\cos \varphi ,$ $x^{2}=r\sin \theta \sin \varphi $\ and $x^{3}=r\cos \theta
, $ followed by \ $\cos \theta =\sin \left( au\theta \left( u\right)
-bv\theta \left( v\right) \right) $ and $y=\varphi $, one obtains potential
1-forms ( Eq.(25)). The YM potentials in Eq.(25) just corresponds to the
curved space generalization of the Wu-Yang ansatz solution\cite{WY}.

\section{Conclusion.}

Customarily YM fields arise as part of a quantum theory inside nuclei. It
carries its own gauge charge and self-interacts with itself. These
properties are fundamentally different from em waves which correspond to the
classical equivalence of photons. In this paper we treat YM waves along with
gravitational waves as classical and consider their collision problem in a
restricted form. This is not to be compared with a Feynman diagram of
quantum chromodynamics. The highly non-linear YM waves acts in this simple
picture parallel to gravitational waves to distort spacetime upon
interaction. The example which we present here is a case that forms Cauchy
horizon ( of Reissner-Nordstrom and KN types) instead of a singularity.
Yasskin's theorem and the Wu-Yang ansatz aided in obtaining this simple
class. Different combinations/collisions which employ the full non-linearity
and non-Abelian character may change this picture completely. This, however,
remains as challenging as ever.

\end{document}